\documentclass[10pt, conference, table]{IEEEtran}
\IEEEoverridecommandlockouts

\usepackage{cite}
\usepackage{amsmath,amssymb,amsfonts}
\usepackage{algorithmic}
\usepackage{graphicx}
\usepackage{textcomp}
\usepackage{hyperref}
\usepackage{enumitem}
\usepackage{booktabs}
\usepackage{comment}

\usepackage{graphicx}
\usepackage{arydshln} 
\usepackage{multirow}
\usepackage{pifont}
\usepackage{cleveref}
\usepackage{array}
\usepackage[normalem]{ulem}
\usepackage{tikz}
\usetikzlibrary{arrows.meta, positioning, calc}
\usepackage{xcolor}
\def\BibTeX{{\rm B\kern-.05em{\sc i\kern-.025em b}\kern-.08em
    T\kern-.1667em\lower.7ex\hbox{E}\kern-.125emX}}
\begin{document}

\title{Towards Intrusion Detection Systems for RPL-based IoT Networks using Foundation Models}

\author{\IEEEauthorblockN{Elias Lunderbye, Sourasekhar Banerjee, Christian Rohner, and Andreas Johnsson}
\IEEEauthorblockA{Uppsala University, Department of Information Technology, Sweden\\ 
}
 }

\maketitle

\begin{abstract}
AI-based intrusion detection systems (IDS) have shown promise in detecting attacks on IoT systems. In this work, we explore the use of foundation models to detect and identify attacks,  
with a specific focus on RPL‑based IoT networks.
We study multiple attack types, attack variations, and network configurations, and provide  insights into the performance of foundation models for attack identification. Specifically, we fine‑tune the MOMENT foundation model for multi‑class attack identification. Our evaluation is based on a dataset containing RPL‑related statistics collected under normal operation and under Blackhole, DIS‑flooding, Worst parent, and Local repair attacks, generated in a Cooja simulation environment. The initial results are promising. The approach achieves attack-detection performance comparable to state‑of‑the‑art methods, while also demonstrating strong performance in distinguishing between different attack types.
\end{abstract}

\begin{IEEEkeywords}
Internet of Things, Intrusion Detection Systems, Foundation Models
\end{IEEEkeywords}

\section{Introduction}

The Internet of Things (IoT) has emerged as an important technological enabler for a wide range of applications critical for society, including healthcare systems, smart power grids, and smart cities \cite{hassan2019current}. These systems rely on large-scale deployments of resource-constrained devices that are often placed in exposed or unattended environments. While such characteristics enable flexible and cost-efficient deployments, they also introduce significant security challenges. In particular, pervasive connectivity combined with limited computational resources and physical exposure makes IoT networks attractive targets for malicious actors. Compromised devices can disrupt routing, degrade network performance, and ultimately threaten the availability and reliability of critical services.

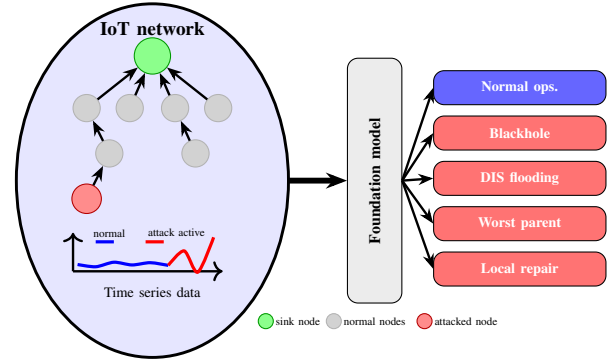
\begin{figure}[t]
\centering
\begin{tikzpicture}[
    scale=0.6, transform shape,
    iot/.style={circle, draw=gray!60, fill=gray!35, minimum size=6mm},
    sink/.style={circle, draw=green!60!black, fill=green!45, minimum size=8mm},
    attacked/.style={circle, draw=red!70!black, fill=red!45, minimum size=6.5mm},
    arrow/.style={-{Stealth[length=2.0mm]}, line width=0.9pt},
    bigarrow/.style={-{Stealth[length=2.8mm]}, line width=2.2pt},
    outbox/.style={draw, rounded corners, minimum width=3.8cm,
                   minimum height=0.75cm, font=\small\bfseries},
    legendtext/.style={font=\scriptsize}
]

\begin{scope}[yshift=-0.45cm]

\draw[draw=black, line width=1.1pt, fill=blue!10]
  (0,0) ellipse (3.0cm and 3.9cm);
\node[font=\large\bfseries] at (0,3.35) {IoT network};

\node[sink] (sink) at (0,2.75) {};
\node[iot] (n1) at (-1.45,1.60) {};
\node[iot] (n2) at (-0.50,1.60) {};
\node[iot] (n3) at (0.50,1.60) {};
\node[iot] (n4) at (1.45,1.60) {};
\node[iot] (n5) at (-0.95,0.6) {};
\node[iot] (n6) at (0.95,0.6) {};
\node[attacked] (att) at (-1.45,-0.4) {};

\draw[arrow] (n1) -- (sink);
\draw[arrow] (n2) -- (sink);
\draw[arrow] (n3) -- (sink);
\draw[arrow] (n4) -- (sink);
\draw[arrow] (n5) -- (n1);
\draw[arrow] (n6) -- (n3);
\draw[arrow] (att) -- (n5);

\begin{scope}[yshift=-0.9cm]

\draw[->, line width=0.9pt] (-1.75,-1.10) -- (1.55,-1.10);
\draw[->, line width=0.9pt] (-1.75,-1.10) -- (-1.75,-0.25);

\draw[blue, line width=1.2pt] plot[smooth] coordinates {
  (-1.65,-0.95) (-1.25,-1.00) (-0.85,-0.90)
  (-0.45,-0.96) (-0.05,-0.91) (0.35,-0.97)
};

\draw[red, line width=1.2pt] plot[smooth] coordinates {
  (0.35,-0.97) (0.70,-0.65) (1.00,-1.12) (1.35,-0.35)
};

\node[font=\small] at (0,-1.55) {Time series data};

\draw[blue, line width=1.2pt] (-1.25,-0.45) -- (-0.85,-0.45);
\node[legendtext, anchor=west] at (-1.42,-0.25) {normal};

\draw[red, line width=1.2pt] (-0.15,-0.45) -- (0.25,-0.45);
\node[legendtext, anchor=west] at (-0.22,-0.25) {attack active};

\end{scope}

\end{scope}

\coordinate (iotout) at (3.0,-0.45);

\node[draw, rounded corners, fill=gray!15,
      minimum width=1.2cm, minimum height=5.4cm,
      font=\bfseries, align=center] (fm) at (4.9,-0.45)
      {\rotatebox{90}{Foundation model}};

\draw[bigarrow] (iotout) -- (fm.west);

\node[outbox, fill=blue!60, text=white] (outN) at (8.1, 1.6) {Normal ops.};

\node[outbox, fill=red!55, text=white] (outBH) at (8.1, 0.6) {Blackhole};
\node[outbox, fill=red!55, text=white] (outDF) at (8.1,-0.4) {DIS flooding};
\node[outbox, fill=red!55, text=white] (outWP) at (8.1,-1.4) {Worst parent};
\node[outbox, fill=red!55, text=white] (outLR) at (8.1,-2.4) {Local repair};

\foreach \x in {outN,outBH,outDF,outWP,outLR}
  \draw[arrow] (fm.east) -- (\x.west);

\node[sink, minimum size=3mm] at (2.5,-3.55) {};
\node[legendtext, anchor=west] at (2.6,-3.55) {sink node};

\node[iot, minimum size=3mm] at (4.0,-3.55) {};
\node[legendtext, anchor=west] at (4.1,-3.55) {normal nodes};

\node[attacked, minimum size=3mm] at (6.0,-3.55) {};
\node[legendtext, anchor=west] at (6.1,-3.55) {attacked node};

\end{tikzpicture}
\caption{A foundation model processes time-series RPL data collected at the sink node in an IoT network, and outputs the network state, i.e., whether the network is in normal operations or under the attack.}
\label{fig:iot-network-fm}
\end{figure}

In this paper, we study intrusion detection and attack identification in low-power and lossy IoT networks operating the Routing Protocol for Low-Power and Lossy Networks (RPL)~\cite{rfc6550}. As illustrated in Figure~\ref{fig:iot-network-fm}, we consider a setting in which a malicious actor compromises a network node and launches attacks to disrupt network operation. We focus on four representative attacks: Blackhole, DIS-flooding, Worst parent, and Local repair \cite{bergqvist2025assessing}. Prior work has demonstrated that approaches based on machine learning can effectively exploit patterns in network telemetry for intrusion detection \cite{al2025comprehensive}. In particular, recurrent models such as Long Short-Term Memory (LSTM) networks have shown improved performance by capturing temporal dependencies in routing statistics \cite{kaveh2025factors}, and the capability of continuous retraining to mitigate deterioration of model performance due to network dynamics \cite{banerjee2026quantifying}. 

In this paper, we build upon the  momentum of foundation models \cite{kolides2023artificial}, that is large pre-trained models that have learned general-purpose representations, for the purpose of intrusion detection. They are attractive for example due to limited need for labeled data in the finetuning process. Unlike prior work, which focused primarily on binary attack detection, this paper studies the effectiveness of using foundation models for not only detecting malicious behavior but also identifying the specific attack type being launched in the RPL-based network.

The main contributions of this paper are twofold: (1) we propose an intrusion detection approach based on pre-trained time-series foundation models, specifically MOMENT \cite{goswami2024moment}, for attack detection and identification in RPL-based IoT networks; and (2) we present an evaluation using data generated in a controlled Cooja \cite{finne2021multi} simulator environment. The results show that pre-trained foundation models are effective in identifying routing attacks in IoT networks.

\section{Problem definition}


In this paper, we investigate the effectiveness of pre-trained foundation models for intrusion detection and attack identification in IoT networks. Specifically, we assess the ability of foundation models to detect whether an RPL-based IoT network is under attack, and if so which attack, using time-series routing data extracted during run time. 

Formally, an intrusion detection system agent $\mathcal{A}$ maintains a time-series foundation model $M$, to exploit temporal patterns in the data, that takes input features $X$ obtained from the network and outputs a label $L_t$ at time $t$, indicating the network state, where $L_t$ belongs to the set of classes
\{Normal, Blackhole, Worst parent, DIS-Flooding, Local repair\}.

The foundation model $M$ is pretrained, and thereafter finetuned using operational network data $X_{t-j,t}$, observed as a multivariate time series over the interval $[t-j,t]$, where $j$ denotes a configurable temporal window. The data $X_{t-j,t}$ consists of RPL routing statistics collected at the IoT sink node, similarly to our previous work \cite{banerjee2026quantifying}. We hypothesize that the considered attacks exhibit temporal patterns that can be exploited for detection and identification. Both $X$ and $L$ evolve over time due to attacker behavior, radio conditions, network load, and environmental effects. 

The objective is to finetune a foundation model, in this paper limited to MOMENT \cite{goswami2024moment}, to learn a mapping
\[
M : X_{t-j,t} \rightarrow L_t
\]
that maximizes attack identification performance and reduces computational resources needed for finetuning.

\section{Scenario}
As in our previous work \cite{kaveh2024impact, kaveh2025factors, bergqvist2025assessing, banerjee2026quantifying}, we consider an RPL-based network \cite{rfc6550} simulated in Cooja \cite{finne2021multi}. RPL organizes IoT nodes into a Destination-Oriented Directed Acyclic Graph (DODAG) rooted at a sink node (green node in Figure \ref{fig:iot-network-fm}). 
The IoT nodes exchange RPL control messages 
to establish and maintain forwarding across the network. As in previous work, we let the sink aggregate control-message statistics from all nodes, to later be used for model training. 

We considered four RPL-based IoT network attacks \cite{bergqvist2025assessing}, briefly described in the following. In the \textbf{\textit{Blackhole (BH)}} attack, a malicious node advertises an artificially low rank to attract traffic, which is then dropped, disrupting packet forwarding. Further, in the \textbf{\textit{DIS-Flooding (DF)}} attack, an attacker repeatedly broadcasts DIS messages, forcing nodes to respond with DIOs and overwhelming the network with control traffic. Furthermore, in the \textbf{\textit{Worst parent (WP)}} attack, a compromised node manipulates rank values, causing neighbors to select suboptimal routes and degrading performance. Finally, in the \textbf{\textit{Local repair (LR)}} attack, an attacker repeatedly triggers false local repairs, increasing routing overhead and delay.

Each of these attacks can manifest in three behavioral variants \cite{kaveh2025factors}: base, on–off, and decreasing. In the base variant, the attack starts suddenly and continues without changing. In the on–off variant, the attacker switches between normal and malicious behavior according to a predefined pattern. Finally,  gradual change is similar to the base variant, but instead of an abrupt start, gradually changes the attack intensity.

The considered networks consist of 20 IoT nodes organized in tree topologies with varying depth. For each scenario, the simulation is divided into two phases: an initial period under normal conditions followed by a period under attack.


\section{Approach}
\subsection{Foundation models for IDS}
MOMENT is a family of open time-series foundation models designed for general-purpose time-series representation learning and fine-tuning~[8]. Architecturally, MOMENT follows an encoder-only transformer design that operates on fixed-length contexts and uses patching to convert the input sequence into a sequence of tokens. There are three different configurations: small, base, and large. 

In this paper, we use the small variant, where MOMENT expects a context length of 512 timesteps. In the current configuration, the internal patch tokenizer uses patch length 8 and patch stride 8, which provides a compact tokenized representation of each multivariate window before it is processed by the pretrained encoder. In the notation of the problem statement, this window corresponds to $X_{t-j,t}$, that is the observed multivariate features over the interval $[t-j,t]$. This makes MOMENT suitable for the learning problem $M : X_{t-j,t} \rightarrow L_t$, where the model uses the windowed input to predict the network-state label at time $t$.

Compared to time-series foundation models such as MOIRAI~[11] and Chronos~[13], tailored for generative forecasting, MOMENT is attractive in our setting because its encoder-based representation pipeline can be reused directly for window-level classification. This better matches the IDS task considered, where the objective is to map the observed routing statistics in $X_{t-j,t}$ to the label $L_t$ rather than to generate future values. Further, the patch-based input representation, where each window $X_{t-j,t}$ is divided into short ordered temporal segments before encoding, fits naturally with the causal temporal windowing used, where observations from the interval $[t-j,t]$ are used to predict the label at time $t$.

\begin{figure}[t]
\centering
\begin{tikzpicture}[
    scale=0.55, transform shape,
    box/.style={draw, rounded corners, line width=1.0pt,
                minimum width=7.4cm, minimum height=0.9cm,
                align=center, fill=gray!10},
    data/.style={draw, rounded corners, line width=1.0pt,
                 minimum width=7.4cm, minimum height=0.9cm,
                 align=center, fill=blue!12},
    model/.style={draw, rounded corners, line width=1.0pt,
                  minimum width=7.4cm, minimum height=0.9cm,
                  align=center, fill=gray!15, font=\bfseries},
    arrow/.style={-{Stealth[length=2.1mm]}, line width=0.9pt},
    small/.style={font=\small}
]

\node[data] (sim) at (0,0)
{Step 1: RPL attack simulations\\
\small IoT-Attacks-IDS};

\node[small] at (0,-0.75)
{$X$: time-series \quad $L$: attack labels};

\node[box] (split) at (0,-1.8)
{Step 2: Run-level split\\
\small Train / Validation / Test};

\node[box] (window) at (0,-3.3)
{Step 3: Causal sliding windows\\
\small Majority-based labeling};

\node[box] (moment) at (0,-4.8)
{Step 4: MOMENT (classification mode)};

\node[small] at (0,-5.6)
{Frozen backbone, finetuned head};

\node[box] (eval) at (0,-6.8)
{Step 5: Evaluation\\
\small F1 score metrics\\
\small Normal / Transition / Attack windows};

\draw[arrow] (sim) -- (split);
\draw[arrow] (split) -- (window);
\draw[arrow] (window) -- (moment);
\draw[arrow] (moment) -- (eval);

\end{tikzpicture}
\caption{Evaluation pipeline for the MOMENT-based IDS.}
\label{fig:evaluation-pipeline}
\end{figure}
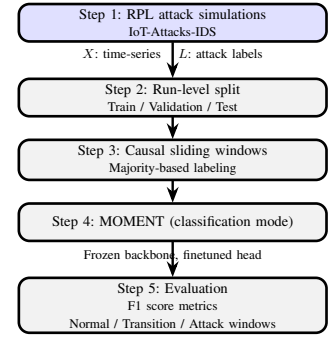

For intrusion detection and attack identification, the key property of MOMENT is that it provides pretrained temporal representations that can be reused with limited labeled data. In our setting, the model is adapted in classification mode by training a lightweight classification head on top of the pretrained encoder so that representations extracted from $X_{t-j,t}$ can be mapped to $L_t$. In the reported baseline, the pretrained patch embedding and transformer encoder are kept frozen, and only the final linear classification head is updated. Concretely, this corresponds to training \texttt{head.linear.weight} and \texttt{head.linear.bias}, that is 23,045 trainable parameters out of 35,360,453 total parameters, or approximately 0.065\% of the model. This restricted fine-tuning scope reduces memory usage and adaptation cost, which is important for practical IDS deployment when labeled attack data are limited.

\begin{figure*}[t]
    \centering
    \includegraphics[width=0.3\textwidth]{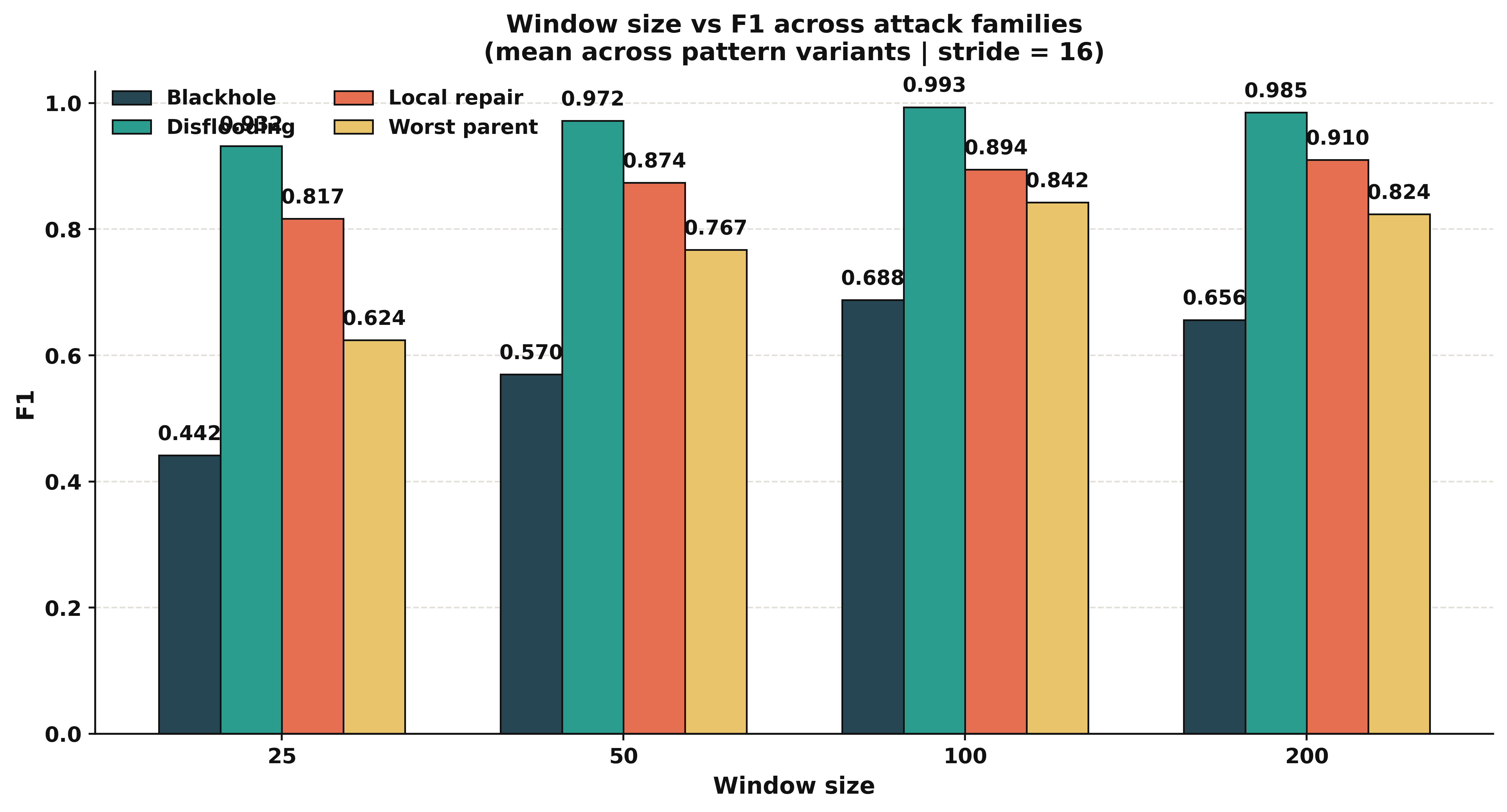}
    \includegraphics[width=0.3\textwidth]{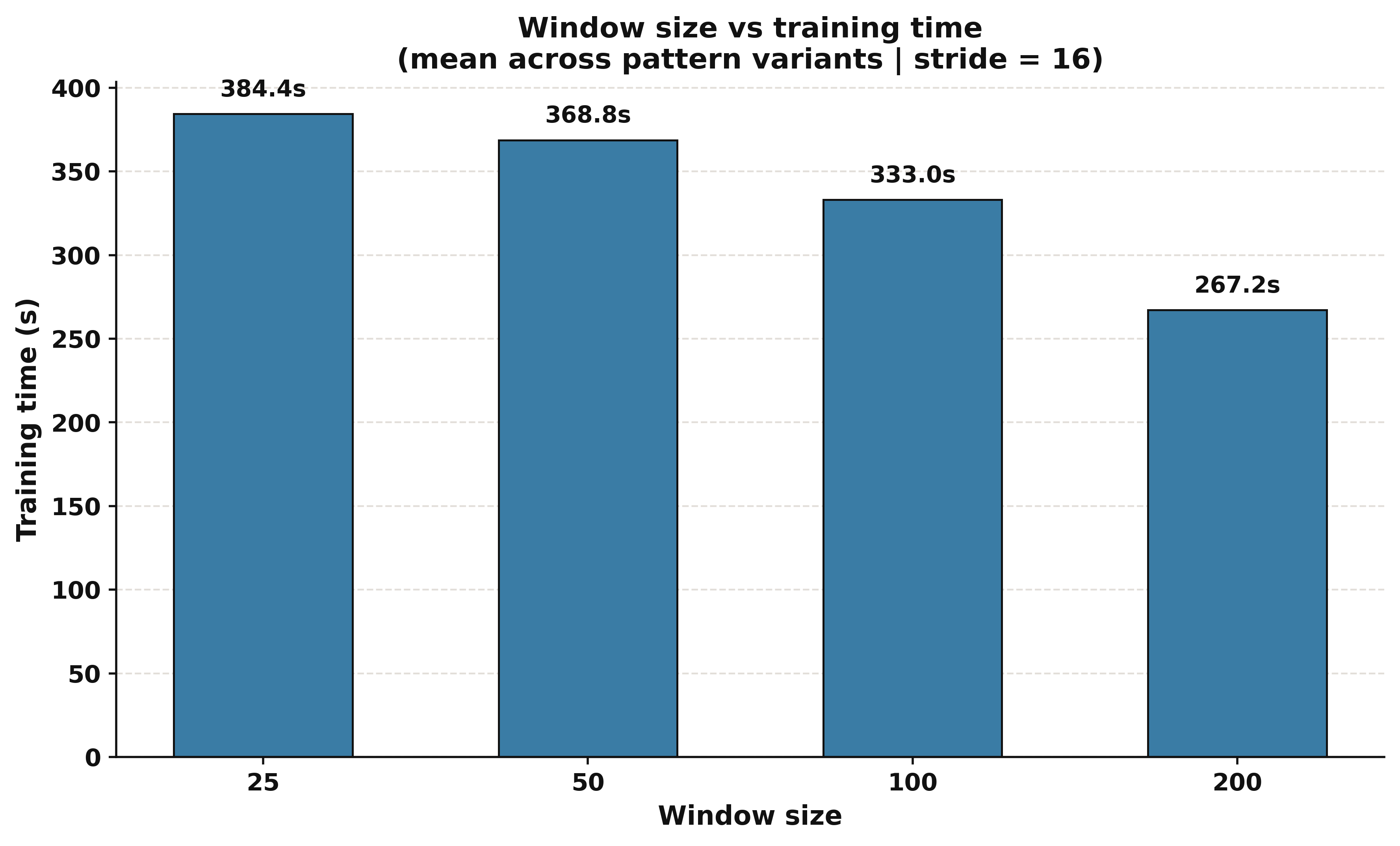}
    \includegraphics[width=0.3\textwidth]{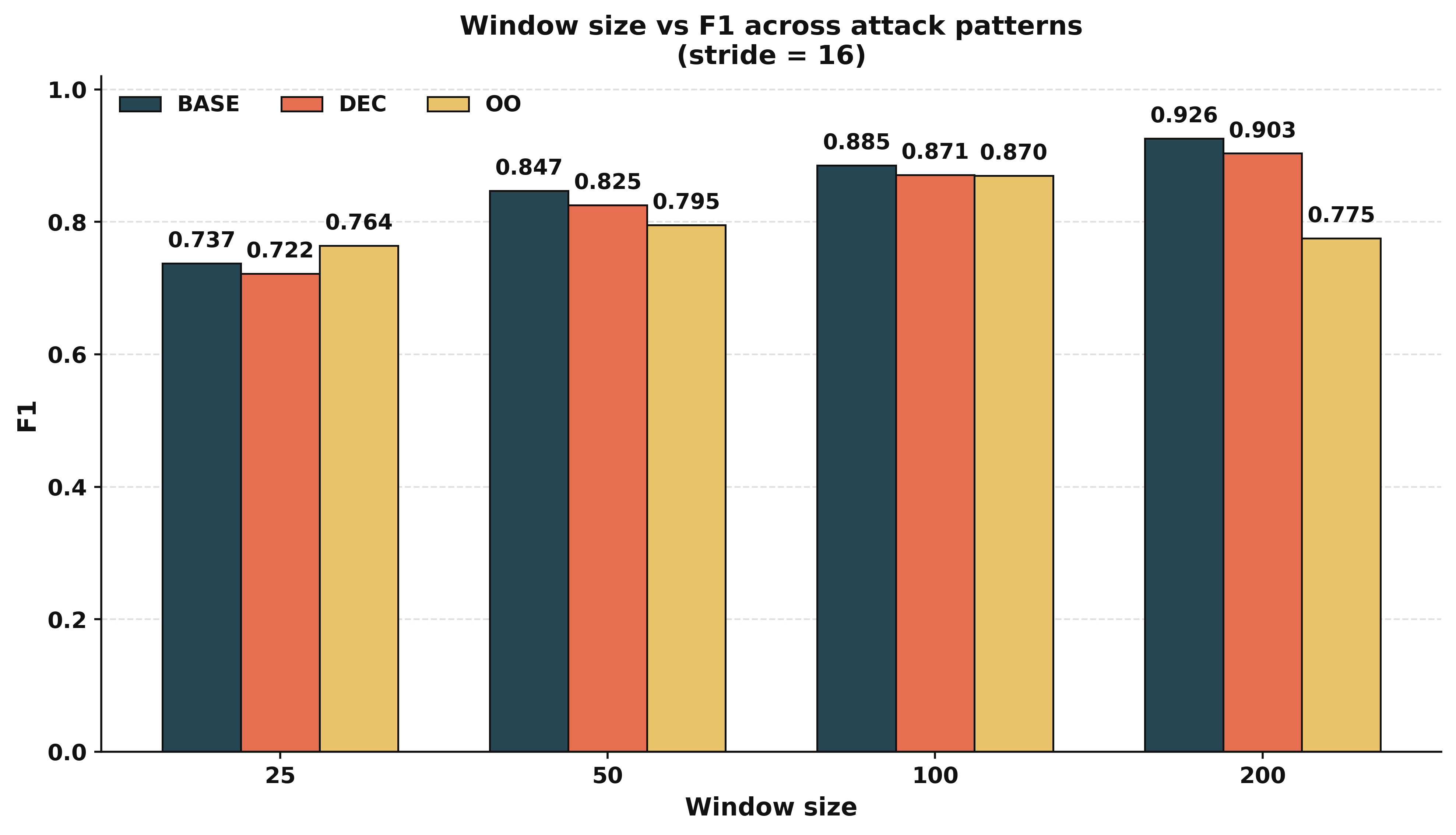}
    \caption{Effect of window size (stride=16) on MOMENT, showing F1 across attack patterns and attack families together with the corresponding training time.}
     \label{fig:window_effects}
\end{figure*}

\begin{figure*}[t]
    \centering
    \includegraphics[width=0.3\textwidth]{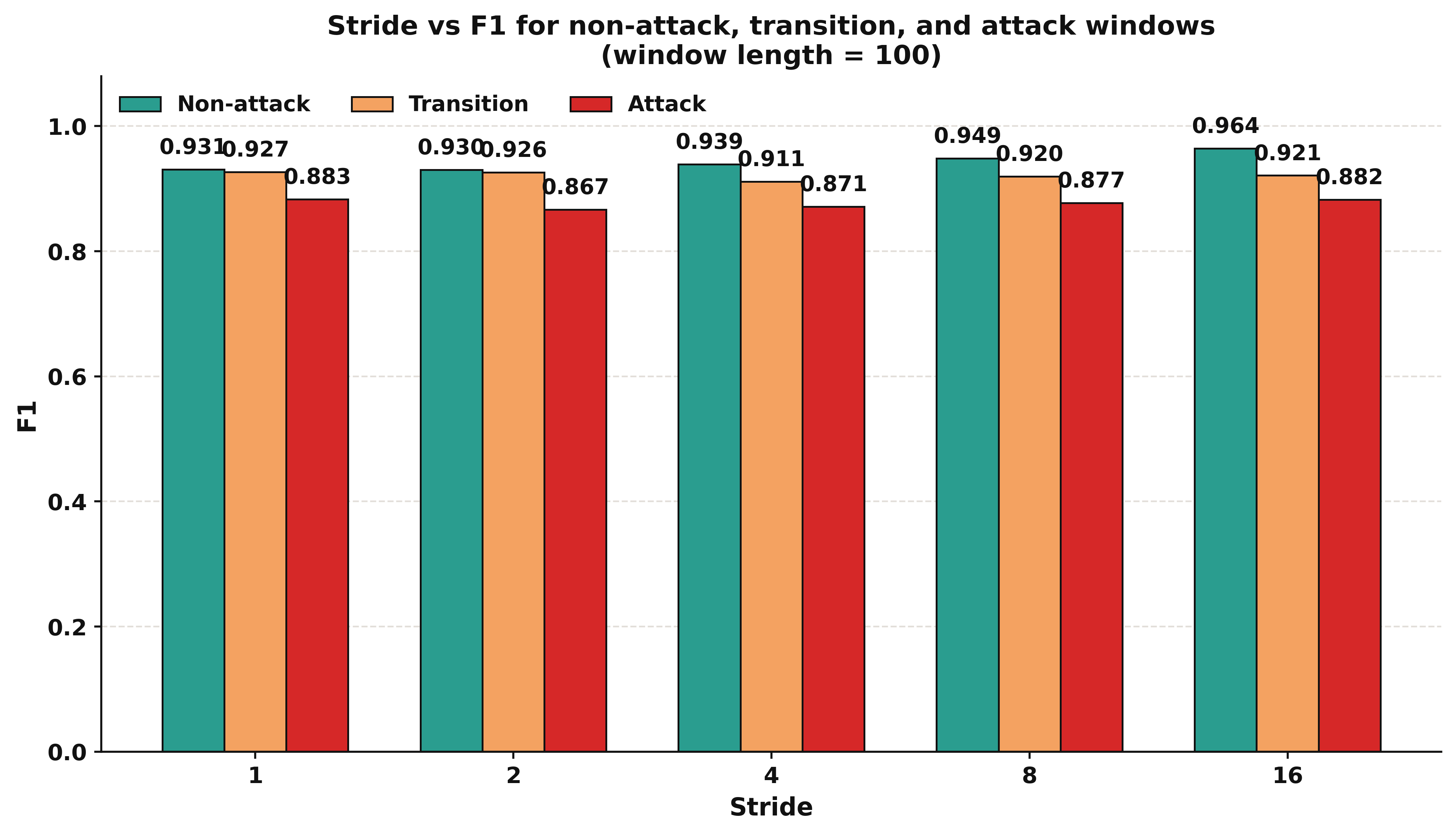}
    \includegraphics[width=0.3\textwidth]{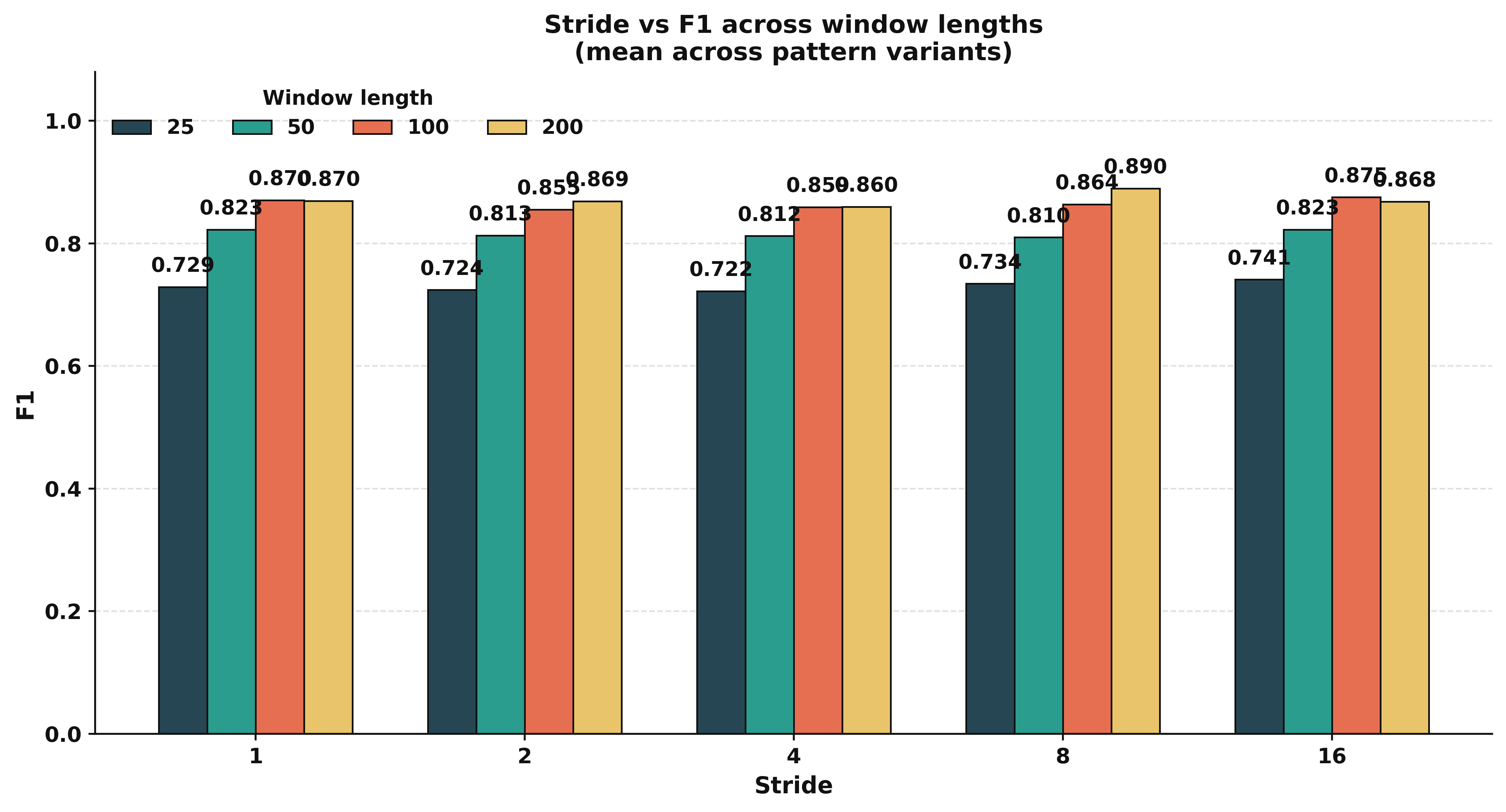}
    \includegraphics[width=0.3\textwidth]{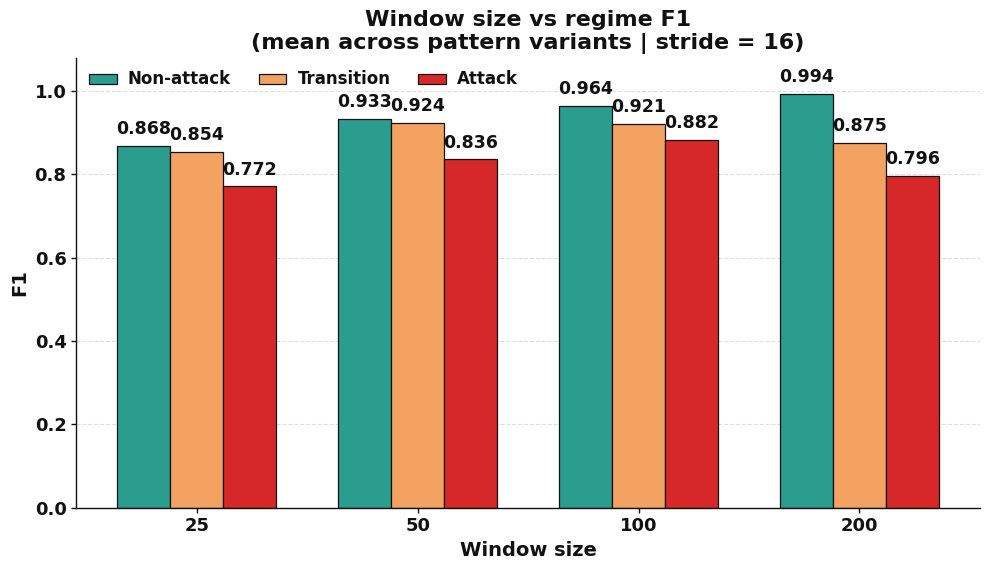}
    \caption{Effect of window size and stride on MOMENT performance, showing F1 across temporal regimes and window sizes.}
    \label{fig:stride_effects}
\end{figure*}

\subsection{Evaluation setting}

The evaluation, and the corresponding pipeline as depicted in Figure~\ref{fig:evaluation-pipeline}, uses multivariate time-series data generated from simulated RPL-based IoT attack scenarios from the IoT-Attacks-IDS repository \cite{iot_attacks_ids}. The repository provides the experimental scenarios together with a pipeline that converts simulation outputs into structured measurements.

In the RPL-attack simulation, at discrete time intervals \( t_1, t_2, \ldots, t_k \), network data is collected at the sink node, including control message exchanges (DIS, DIO, DAO), routing updates (rank), and traffic statistics, capturing both normal operation and attack conditions. The data is then transferred to a server and converted into time-series stored in CSV files. The extracted feature sequences correspond to $X$, while the associated attack annotations define the label sequence $L$. Accordingly, $X$ represents the observed protocol-related time-series features and $L_t$ the target class at time $t$, indicating normal behavior or one of the considered attack families.

For data management, each simulation run is treated as one multivariate time series and partitioned into training, validation, and test sets to avoid leakage between windows from the same run, as illustrated in step 2 of Figure~\ref{fig:evaluation-pipeline}. The training split is used to compute feature standardization statistics, after which each run is segmented into causal sliding windows $X_{t-j,t}$. A window is labeled as normal when it contains no attack activity; otherwise, it is assigned the corresponding attack class once the fraction of attack-labeled samples in $L$ exceeds the selected majority threshold. This provides a consistent representation of non-attack, transition, and attack periods without introducing low-level file handling details, corresponding to step 3 in the figure.

In step 4, MOMENT is used as a pretrained time-series foundation model in classification mode. Each window $X_{t-j,t}$ is represented as a multichannel sequence and mapped to the model input format through padding and masking when required. The classification head is fine-tuned while the pretrained backbone is kept fixed, allowing the model to adapt to five-class intrusion identification with limited trainable parameters. 

Finally, in step 5, performance is evaluated at the window level using  macro-F1, the unweighted mean of per-class F1 (the precision--recall harmonic mean), so all five potentially imbalanced classes contribute equally. To further assess temporal behavior, we also report results separately for non-attack, transition, and full-attack windows. 
We also track computational cost to assess 
deployment feasibility. 

\section{Results}
The evaluation results are shown in Figures~\ref{fig:window_effects} and \ref{fig:stride_effects}. An initial sensitivity analysis shows that window size has a stronger effect on MOMENT performance than stride. In the left panel of Figure~\ref{fig:window_effects}, longer windows improve performance for most attack families up to a window size of 100. \textit{DIS-flooding} is consistently the easiest family to identify and \textit{Blackhole} the hardest, while \textit{Worst parent} and \textit{Blackhole} decline slightly at 200, indicating diminishing returns beyond 100.

The middle panel of Figure~\ref{fig:window_effects} shows that larger windows are not more expensive in this setup. Instead, average training time decreases from 384.4\,s at length 25 to 267.2\,s at 200, likely because fewer windows are generated from each sequence. The right panel shows the same overall pattern at the attack-variant level: mean F1 rises from 0.741 at length 25 to 0.875 at 100, then decreases slightly at 200. The \textit{base} and \textit{dec} variants continue to improve with longer windows, whereas the \textit{oo} variant peaks at 100 and drops at 200. Overall, Figure~\ref{fig:window_effects} suggests that a window size of 100 provides the best trade-off between accuracy and cost.

Figure~\ref{fig:stride_effects} shows that stride has a weaker effect than window size. In the left panel, for window size 100, regime-level F1 varies only modestly across strides 1--16, with non-attack windows improving slightly as stride increases. The middle panel confirms this at the aggregate level, where mean F1 changes little across strides for a fixed window size. In the right panel, window size again dominates: non-attack performance improves steadily with longer windows, while transition and attack windows peak around 50--100 and then decline at 200. Based on these results, we use a window size of 100 and stride 16 in the subsequent analysis, since this configuration provides good balance between attack-phase performance, temporal robustness, and computational efficiency.

Further analysis shows that most errors occur in full-attack windows rather than during attack onset. For the main setting (window 100, stride 16, majority threshold 0.30), transition windows are rarely confused with normal behaviour, while the dominant remaining error is confusion between \textit{Local repair} and \textit{Worst parent}. 
Blackhole resembles  normal traffic in sink-level features, thus making it the hardest attack to detect. 

For a representative configuration with window length 100 and majority threshold 0.30, training required 311.48\,s and inference averaged 7.80\,ms per window on the evaluation platform. Although a dedicated sink-node benchmark is left for future work, these numbers indicate that the proposed pipeline is compatible with near-real-time analysis at the sink node.

Finally, the closest baseline is the LSTM-based IDS in~[5], which considers the same RPL/Cooja sink-node setting but only for binary attack detection. The present results extend that baseline to attack identification: \textit{Blackhole} remains the hardest case and \textit{DIS-flooding} the easiest, but the task is expanded to five-class family prediction with temporal-regime analysis.

\section{Related works}
Recent advances in AI have introduced foundation models for time‑series analysis, aiming to learn general‑purpose temporal representations. MOMENT \cite{goswami2024moment}, as used in this paper, is a time‑series foundation model pre‑trained on diverse datasets, supporting a variety of use cases. Other notable time-series foundation models include MOIRAI \cite{liu2024moirai}, TimeGPT \cite{garza2023timegpt}, Chronos \cite{ansari2024chronos}, and Tiny Time Mixers (TTM) \cite{ekambaram2024tiny}.
Recently, foundation models have also been applied to intrusion detection. In \cite{zhou2026traffic}, Traffic‑MoE is introduced as a foundation model for network traffic analysis and anomaly detection, and in \cite{garcia2025foundation} the authors evaluate TabPFN for enabling intrusion detection in IoT networks. In these cases, a pretrained model is used for IDS, but target general traffic or tabular IoT data rather than RPL time-series data.
Conversely, RPL IDS studies such as \cite{kaveh2024impact, kaveh2025factors, bergqvist2025assessing, banerjee2026quantifying} address LSTM model generalizability, but not reusable foundation models. Thus, the application of time-series foundation models to  attack detection in RPL-based IoT networks remains largely unexplored. 

\section{Conclusions}
In this paper, we proposed a time-series foundation model approach for intrusion detection and attack identification in RPL-based IoT networks. Specifically, we finetuned the MOMENT foundation model to identify four representative routing attacks: Blackhole, DIS-flooding, Worst parent, and Local repair. Our  evaluation shows that the proposed approach achieves attack detection performance comparable to state-of-the-art methods, while also providing promising multi-class attack identification results for all attacks except the Blackhole attack.
We further observe that the choice of temporal window size has a noticeable impact on both training time and classification performance. In particular, a window size of 100 provides a good trade-off between computational cost and F1 score, making it a practical choice for the considered setting.
As future work, we will explore additional foundation models to assess their general applicability for intrusion detection in IoT networks. We also aim to improve performance on challenging attack types and evaluate the approach across broader scenarios.


\section*{Acknowledgment}
This research has been supported by the Swedish Govern-
mental Agency for Innovation Systems (VINNOVA) through
the project Robust IoT Security: Intrusion Detection Lever-
aging Contributions from Multiple Systems (2023-02982),
as well as the Swedish Civil Contingencies Agency (MSB)
through the Robust IoT project (2018-12526). 

\bibliographystyle{IEEEtran}
\bibliography{bibtex/bib/bibliography} 





\end{document}